\documentclass[pdflatex,sn-mathphys]{sn-jnl}
\jyear{2024}
\theoremstyle{thmstyleone}

\theoremstyle{thmstyletwo}

\theoremstyle{thmstylethree}

\usepackage{amsmath,amssymb,amsfonts}
\usepackage{mathrsfs}
\usepackage{algorithm}
\usepackage{algorithmicx}
\usepackage{textcomp}
\usepackage{xcolor}
\usepackage{url}
\usepackage{multirow}
\usepackage{tabularx}
\usepackage{lipsum}
\usepackage{comment}
\usepackage{subcaption}
\usepackage{graphicx}
\usepackage{colortbl}
\usepackage{geometry}

\begin{document}
\title[Preliminary evidence]{Human Brain Exhibits Distinct Patterns When Listening to Fake Versus Real Audio: Preliminary Evidence}

\author[1]{\fnm{Mahsa} \sur{Salehi}} \email{mahsa.salehi@monash.edu}
\author[2]{\fnm{Kalin} \sur{Stefanov}} \email{kalin.stefanov@monash.edu}
\author[1]{\fnm{Ehsan} \sur{Shareghi}} \email{ehsan.shareghi@monash.edu}

\affil[1]{\orgdiv{Department of Data Science and Artificial Intelligence}, \orgname{Monash University}, \orgaddress{\city{Melbourne}, \state{Victoria}, \country{Australia}}}
\affil[2]{\orgdiv{Department of Human-Centred Computing}, \orgname{Monash University}, \orgaddress{\city{Melbourne}, \state{Victoria}, \country{Australia}}}

\abstract{
In this paper we study the variations in human brain activity when listening to real and fake audio. Our preliminary results suggest that the representations learned by a state-of-the-art deepfake audio detection algorithm, do not exhibit clear distinct patterns between real and fake audio. In contrast, human brain activity, as measured by EEG, displays distinct patterns when individuals are exposed to fake versus real audio. This preliminary evidence enables future research directions in areas such as deepfake audio detection. 
}


\maketitle

\section{Introduction} 
Increasingly powerful deep learning algorithms accompanied by the rapid advances in computing power have enabled the generation of hyper-realistic synthetic media commonly referred to as \emph{deepfakes}.
Deepfakes include all forms of digital content - video, text, images, and audio - that have been either manipulated or created from scratch using deep learning algorithms to primarily mislead, deceive or influence audiences~\cite{UNI}.
In addition, access to deepfake generation technologies has become widespread, and the technologies are easy to use - deepfake generation technologies have been democratised~\cite{Weaponised}.
However, deepfake generation technologies provide new tools to cybercriminals with broad impact in cyber-enabled crime.
For example, audio generation and in particular voice cloning (creating a synthesised voice from a small audio sample of an authentic voice) can be used in sophisticated attacks.
In 2019, criminals used voice cloning technology to impersonate an executive's voice in the first reported use of deepfakes in a cybercrime operation, deceiving the CEO of a United Kingdom energy firm into transferring \$240,000~\cite{Wallstreet}.
In 2020, cybercriminals cloned the voice of a company director in the United Arab Emirates to steal as much as \$35 million~\cite{35Mil}.

In the past 5 years, we have witnessed a dramatic increase in the number of publications per year that address issues related to deepfake content detection - from less than $250$ publications in $2018$ to more than $3250$ in $2022$~\footnote{app.dimensions.ai}.
In $2022$, at the IEEE International Conference on Acoustics, Speech and Signal Processing, the first Audio Deep Synthesis Detection Challenge was launched to motivate the development of deepfake audio detection algorithms~\cite{ADD2022}.
The body of research on deepfake detection is driven by datasets created with different deepfake generation techniques.
Consequently, most deepfake detection methods are designed, developed, and evaluated on a subset of manipulations, and there is no guarantee that they will generalise to unseen classes of manipulations.
The nature of voice cloning-enabled crime (\emph{i.e.,} the fact that victims have expectations for the voice qualities from prior experiences) motivates formulating the problem of cloned voice detection as anomaly detection.
Anomaly detection is a data mining task and is the process of finding rare events (in this context, the cloned voice) in a dataset which are significantly different from normal events (in this context, the real voice).
Like deepfake detection, one of the challenges in detecting anomalies is that there are many types of anomalies which makes the process of detecting anomalies hard and most existing anomaly detection methods are likely to fail in generalising to novel anomalous events.

Despite the many remarkable successes of machine learning, there is still a gap between human and machine learning.
Evidence for this gap is that patterns of errors made by today's machine learning algorithms are different from the patterns of humans performing the same tasks.
While the human brain has been a source of inspiration for machine learning, little effort has been made to directly use data collected from brain activities as a guide for machine learning algorithms~\cite{fong2018}.
To address the generalisation problem outlined above, this paper proposes anomaly detectors (\emph{i.e.,} cloned voice detectors) that are guided by human brain activities (\emph{i.e., EEG}).
The hypothesis is that the neural activity carries information for the concept of anomaly (\emph{i.e.,} ``meta level'') which is agnostic to the exact type of the anomaly (\emph{i.e.,} ``surface level'').
This hypothesis is supported by a recent study suggesting that we can reliably decode deepfake content using human neural activity~\cite{moshel2022you}.

\section{Method}

\subsection{Problem Formulation}
Given a labeled audio time series dataset $A$, and an equivalent labeled EEG time series dataset $E$ of a human subject listening to $A$, where a label of each sentence in $A$ belong to a set $labels=\{real,fake\}$, the problem is to train a classifier on $E$ to classify $A$ using $labels$. We formulate this as a time series classification problem.

\subsection{Data Collection}

\paragraph{Real Audio Data.}
We collected audio data with ethics clearance for deepfake audio production. We hired 20 native English-speaking actors (10 males, 10 females) between the age of 20-35. We prepared 10 conversational scripts with controlled topics for producing 10 single-turn (\emph{i.e.,} monologue) audios. Each script roughly took 3 minutes to read. We required to collect the data ourselves (as opposed to using publicly available alternatives) to have full control over topics, prosodic features and have the ethical consents to release this unique dataset to the research community.

\paragraph{Deepfake Audio Generation.}
To generate high-quality, realistic fake audio, we employed two state-of-the-art speech generation methods, VITS~\cite{kimConditional2021} and YourTTS~\cite{casanovaYourTTS2022}. VITS is a few-shot speech generation method capable of producing very realistic fake voice through fine-tuning on real data. After fine-tuning, the speaker-specific VITS models receive text inputs and generate fake audio based on the styles of the target speaker. YourTTS is a zero-shot speech generation method that can mimic any speaker's style without fine-tuning. The YourTTS model is pre-trained on multiple datasets to ensure generalizability. The YourTTS model takes text and reference audio as input and generates fake audio with a similar style to the reference audio.

To analyze the impact of different quality settings of the deepfake voices, we defined three quality configurations:

\begin{itemize}
\item \textbf{High}: Fine-tuning VITS on all recorded real speech, then generating target audio included in the training set.
\item \textbf{Medium}: Fine-tuning VITS on the first minute of recorded real speech, then generating target audio that might not be included in the training set.
\item \textbf{Low}: Generating fake audio using YourTTS~\cite{casanovaYourTTS2022} without fine-tuning.
\end{itemize}

Inspired by~\cite{caiYou2022,CAI2023103818,zhangInitial2021}, we replaced a real segment in the real audio with a fake segment to analyze how humans respond to deepfake audio. We randomly chose the position for insertion from three options: 1) After one minute of real audio; 2) In the middle of the second minute of real audio; and 3) At the end of the real audio. Additionally, we only considered replacing segments with more than five words. Subsequently, we applied the speech generation methods to generate three corresponding fake audios of different quality. In the final step, following~\cite{caiYou2022}, we normalized the loudness of fake and real segments to ensure smooth transitions between them. Finally, we concatenated all segments to build the final fake audio samples.

\begin{figure}
    \centering
    \includegraphics[width=0.6\textwidth]{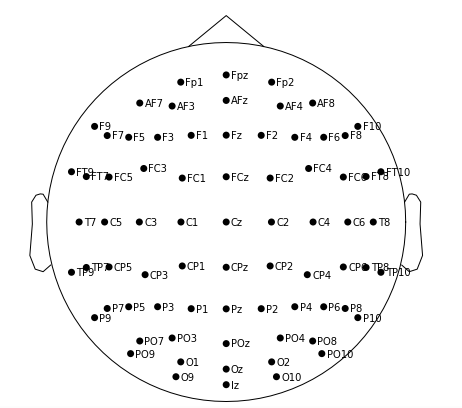}
    \caption{Electrode layout of 64-channel Easycap}
    \label{fig:easycap}
\end{figure}

\paragraph{EEG Data.} 
We collected brain EEG from the participants who listen to mixed audio (audio that contains both read and fake) and are asked to detect deepfake segments and explain their rationale. The data collection sessions took place in the Monash Biomedical Imaging (MBI) center in Monash University. Participants were asked to close their eyes throughout the experiment including the very first 3 minutes baseline recording at the beginning of the experiments. We have used a 64-channel Easycap EEG recording caps following the 10-20 system. Fig.~\ref{fig:easycap} the electrode layout. The recording frequency of the EEG caps is 5000 Hz.

We hired 2 English-speaking participants to record EEG and the corresponding reaction (press a button) when a deepfake was detected. The participants will be instructed not to be concerned with fact checking of statements. To the best of our knowledge there is no public EEG dataset thare are paired with mixed audio like ours. Our dataset will be unique for its controlled construction, scale, and more importantly for containing information along 3 axes: original and deepfake audios, EEG, and participant labeling. This will not only serve the basis of our work but will also be supporting further research in this underexplored space in ML, psychology and neurosciences. Note there are two possibilities - one is that the humans can detect the deepfakes, in which case the EEG can help understand how they detect them. Another is that the EEG shows different brain activity when confronted with deepfakes, but the person involved does not identify the speech as deepfake consciously. 

\subsection{Audio Preprocessing}
Given that the raw audio recordings were noisy, we performed pre-processing to normalize the audio. Firstly, we manually edited the raw audio to remove repetitions, mistakes, and unrelated speech, ensuring the content was clean for input into the speech generation system. Subsequently, we partitioned the audio into several segments based on the transcripts to create clean audio-text pairs for each speaker.

\subsection{EEG Preprocessing}
In this section we explain standard EEG preprocessing steps~\cite{bigdely2015prep} we took to prepare EEG data for analysis.

\begin{figure}
    \centering
    \includegraphics[width=0.6\textwidth]{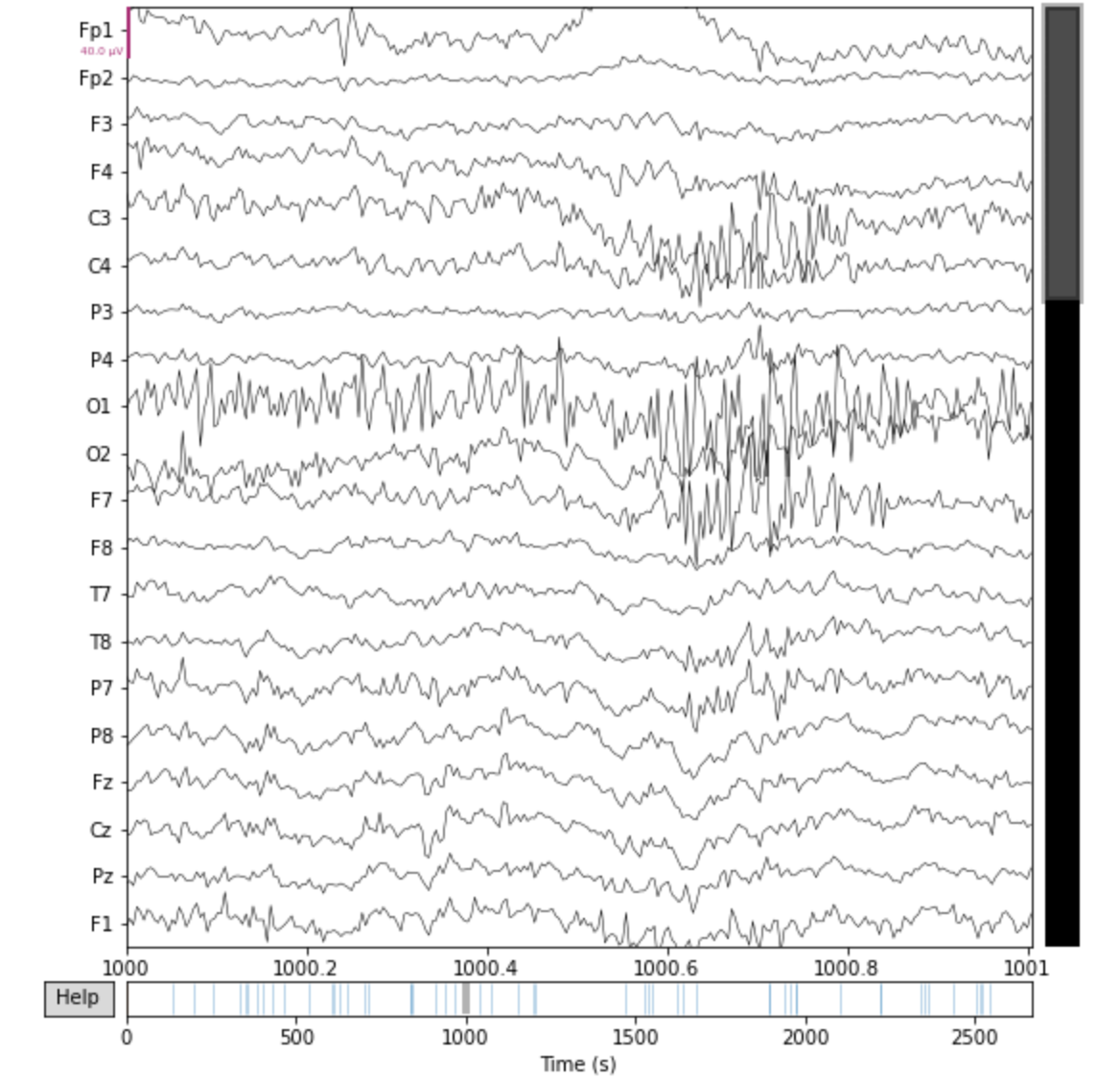}
    \caption{1-second Raw EEG data of Subject 1}
    \label{fig:rawEEG}
\end{figure}

\paragraph{Baseline correction.} We collected 3 minutes of EEG recording without any stimuli (playing audio) at the beginning of our EEG data collection experiment as a baseline. In order to only analysis the effect of our stimuli, we correct our recordings using this baseline data. We used mean baseline correction as suggested in the literature~\cite{bigdely2015prep} which is basically subtracting the baseline mean from the raw EEG. A sample snapshot of our raw EEG is shown in Fig.~\ref{fig:rawEEG}. As can be seen the raw EEG is noisy, containing different types of artifacts such as muscle movement. A standard technique used to remove artifacts is Independent Component Analysis (ICA).

\paragraph{Filtering to remove slow drifts.} We filter the data to remove low-frequency drifts, which can negatively affect the quality of the ICA fit. The slow drifts are problematic because they reduce the independence of the assumed-to-be-independent sources (e.g., during a slow upward drift, the neural, heartbeat, blink, and other muscular sources will all tend to have higher values), making it harder for the ICA algorithm to find an accurate solution. To this end we apply a 0.5-80 Hz bandpass filter.

\paragraph{Re-referencing.} We apply common average re-reference to ensure that all channels contribute equally towards the reduction in channel amplitude, minimizing bias towards specific brain regions. In order to obtain the voltage signal of each channel we use the average of the left and right mastoid.

\begin{figure}
    \centering
    \includegraphics[width=0.6\textwidth]{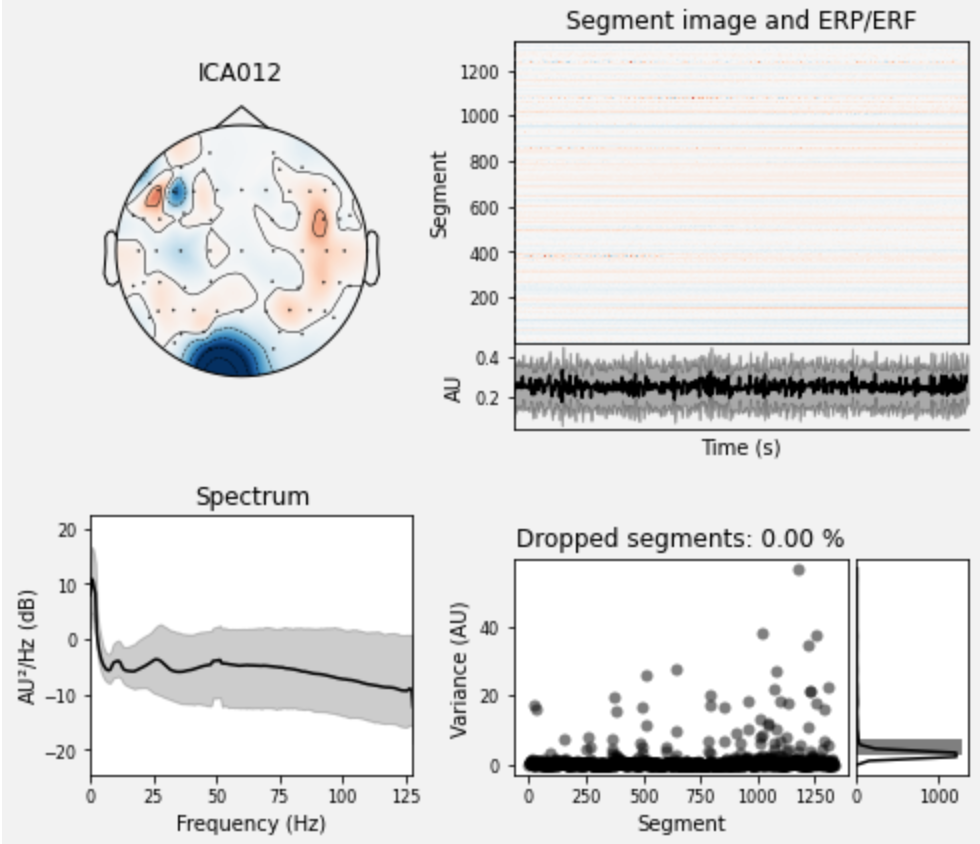}
    \caption{An example of a muscle artifact component in Subject 1}
    \label{fig:ica}
\end{figure}

\paragraph{Artifact removal.} We applied ICA with 20 components to remove potential artifacts in EEG. After decomposition, we then applied widely used automatic ICA component labeling algorithm called ICLabel~\cite{pion2019iclabel}, which will assign a probability value for each component being one of the following categories: muscle artifact, heart beat, line noise, channel noise, eye blink and other. Note our data does not contain eye blink artifact as the participants were asked to close their eyes. ICLabel produces predicted probability values for each of these categories. We visually inspected all the components and the non-artifact components (``Brain'' and ``Other'' categories) were reasonable Fig.~\ref{fig:ica} shows an example of a muscle category that was removed. ``Other'' is a catch-all that for non-classifiable components. We will stay on the side of caution and assume we cannot blindly remove these. After removing the artifact components we reconstructed the EEG back to its original space. We then resampled our EEG data to 256 Hz to be compatible with our audio recordings. 

\paragraph{Segmentation.} We segment the processed EEG recording into segments of time series of 0.5 second windows. Each time series window overlaps 50\% with its before and after segment. Since the frequency rate is 256 Hz, the number of data points in each time series is 128. We denote a EEG segment as $E_i=\{e_1,e_2,\ldots,e_L\}$ where $L=128$ is the length of the time series and $e_j\in \mathbb{R}^{d}$, where $d=64$ is the number of channels. Finally, the segments in which the silence audio is played are removed. This results in a a total of 7548 samples of time series ($E_i$s). 

\begin{table}[h]
\centering
\setlength{\tabcolsep}{5pt}
\setlength\extrarowheight{2pt}
\caption{Description of dataset used in our experiments.}
\begin{tabular}{l|ccccc} 
\hline
Dataset & \multicolumn{1}{l}{\#Train} & \multicolumn{1}{l}{\# Test}  & \multicolumn{1}{l}{Length} & \multicolumn{1}{l}{\# Channel} & \multicolumn{1}{l}{\# Class} \\ \hline \hline
EEG- Subject 1   & 6038 & 1510 & 128  & 64  & 2 \\

\hline
\end{tabular}
\label{Tab:Data}
\vspace{-0.25cm}
\end{table}

\section{Experimental Results}
Our experiments are divided to two categories: 1) We apply statistical tools to visualise our Audio and EEG datasets, and 2) We apply time series classification tools to classify fake and real classes using our EEG dataset. We used data from one of our subjects (Subject 1) in this section. The details of this dataset is presented in Table~\ref{Tab:Data}.

\begin{figure}
    \centering
    \includegraphics[width=0.65\textwidth]{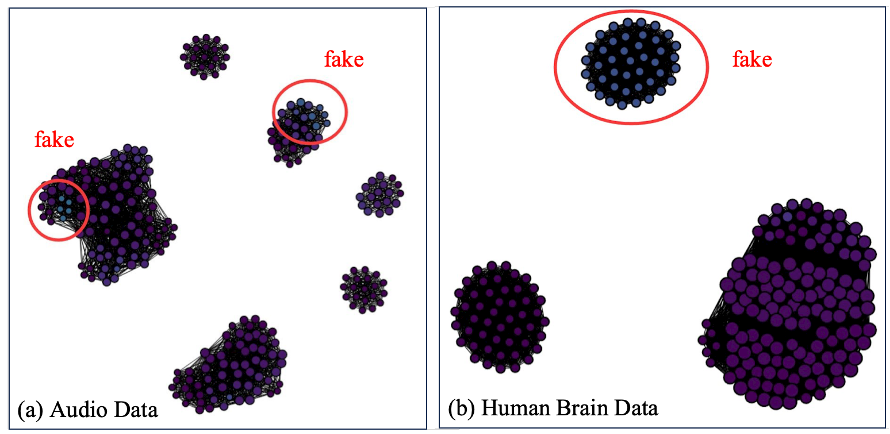}
    \caption{Visualisation of our pilot study}
    \label{fig:pilot}
\end{figure}

\paragraph{Highest Density Regions Visualisation}
We visualize the Highest Density Regions (HDR;\cite{hyndman1996computing}) of the samples of EEG time series ($E_i$s), as well the HDR of the representations of audios from the classification head of the state-of-the-art audio deepfake detection system\footnote{\url{https://huggingface.co/HyperMoon/wav2vec2-base-960h-finetuned-deepfake}} using the Mapper\footnote{\url{https://github.com/scikit-tda/kepler-mapper}} algorithm~\cite{singh2007topological}. The HDR cuts the overall density space to create latent spaces that contain above a threshold probability mass (i.e., minimum ditance $\geq 0.65$ 
). The Mapper's output is represented as a graph in which each component corresponds to a cluster of points that together form a high density region. The connectivity of the graph reflects some topoligical characteristics of the representation space, with darker colors (i.e., dark purples reflecting higher density). 
Blue represents fake and purple represents real. Fig.~\ref{fig:pilot} demonstrates the Mapper output on the audio representations versus EEG representations. The observation highlights a much more clear separation of fake and real signals in the EEG space compared to the audio's. This is an strong indication that EEG's representation of deepfake audio could offer a more discriminative representation of data.

\paragraph{Brain Activity Classification.} In the experiments we use the EEG dataset of $E_i$ samples of subject 1 tp train and test a classifier. We used the state-of-the-art time series classification method called ConvTran~\cite{foumani2024improving} to classify fake and real classes using our EEG dataset. Convtran uses CNN and Transformer to classify samples of time series taking into account the order of data points in each $E_i$ effectively. A two-tower network each using ConvTran is used. The first tower processes the raw data $E_i$, while the second tower processes the FFT of the input, i.e., $FFT(E_i)$. This tower extracts the frequency features including frequency bands such as alpha and gamma. 
The outputs of both towers are concatenated for classification. 

The train and test split is done in two ways. In our first method we selected $E_i$ EEG time series randomly, see Table~\ref{Tab:Data} for the train and test split. The results of In the second experiment we used the first 80\% of the $E_i$s in the data collection session for train and the rest for test. This is a harder task given that the content of the audio was changing throughout the data collection. 
The results are shown in Tables~\ref{Tab:result1} and \ref{Tab:result2}. The results are very promising and suggesting that we can learn the patterns of fake and real only suing EEG data of human listeners listening to real and fake audio.

\begin{table}[h]
\centering
\setlength{\tabcolsep}{5pt}
\setlength\extrarowheight{2pt}
\caption{EEG classification results on Subject 1- random train/test split}
\begin{tabular}{l|ccc} 
\hline
Class & \multicolumn{1}{l}{Precision} & \multicolumn{1}{l}{Recall}  & \multicolumn{1}{l}{F1-score} \\ \hline \hline
Real class  & 0.999 & 0.999 & 0.999  \\
Fake class  & 0.961 & 0.980 & 0.970 \\

\hline
\end{tabular}
\label{Tab:result1}
\vspace{-0.25cm}
\end{table}

\begin{table}[h]
\centering
\setlength{\tabcolsep}{5pt}
\setlength\extrarowheight{2pt}
\caption{EEG classification results on Subject 1- ordered train/test split}
\begin{tabular}{l|ccc} 
\hline
Class & \multicolumn{1}{l}{Precision} & \multicolumn{1}{l}{Recall}  & \multicolumn{1}{l}{F1-score} \\ \hline \hline
Real class  & 0.982 & 0.935 & 0.958  \\
Fake class  & 0.125 & 0.356 & 0.185 \\

\hline
\end{tabular}
\label{Tab:result2}
\vspace{-0.25cm}
\end{table}

\bibliography{Main}

\end{document}